\documentclass[english,aps,prd,%twocolumn,
groupedaddress,amssymb,
showpacs,nofootinbib]{revtex4}

\usepackage{times}
\usepackage{amsmath}
\usepackage{graphicx}

\newcommand{\beq}{\begin{eqnarray}}
\newcommand{\eeq}{\end{eqnarray}}

\makeatletter
\numberwithin{equation}{section}
\numberwithin{figure}{section}

\makeatother

\usepackage{babel}

\begin{document}
\setlength{\unitlength}{1mm}

\title{Multiparticle Form Factors of the Principal Chiral Model At Large N}

\author{Axel \surname{Cort\'es Cubero}}

\email{acortes_cubero@gc.cuny.edu}

\affiliation{ Baruch College, The 
City University of New York, 17 Lexington Avenue, 
New 
York, NY 10010, U.S.A. }

\affiliation{ The Graduate School and University Center, The City University of New York, 365 Fifth Avenue,
New York, NY 10016, U.S.A.}

\begin{abstract}
We study the sigma model with $SU(N)\times SU(N)$ symmetry in $1+1$ dimensions. The two- and four-particle form factors of the Noether current operators are found, by combining the 
integrable-bootstrap method with the large-$N$ expansion.
\end{abstract}

\maketitle

\section{Introduction}

The quantum principal chiral sigma model is completely integrable in one space and one time dimension \cite{Wiegmann}, \cite{Abadalla}. Its action is
\beq
S=\frac{N}{2g_0^2}\int d^2x\,\eta^{\mu\nu} {\rm Tr}\partial_\mu U(x)^\dag\partial_\mu U(x),
\eeq
where $U(x)\in SU(N)$, $\mu,\nu=0,1$,  and where $\eta^{\mu\nu}$ is the Minkowski metric, $\eta^{00}=1, \eta^{11}=-1, \eta^{01}=\eta^{10}=0$. The action is invariant under the global transformation $U(x)\to V_LU(x)V_R$, for $V_L,V_R\in SU(N)$. 
The model is asymptotically free and has a mass gap $m$.
There are two Noether currents, 
\beq
j_\mu^L(x)^c_a=\frac{-iN}{2g_0^2}\partial_\mu U_{ab}(x) U^{*bc}(x),\,\,\, j_\mu^R(x)^d_b=\frac{-iN}{2g_0^2}U^{*da}(x)\partial_\mu U_{ab}(x),\label{currents}
\eeq
where where $a,b=1,...,N$., associated with the symmetries $U\to V_LU$ and $U\to UV_R$ respectively.

In this paper, we calculate the two- and four-excitation form factors of the current operators using a large-$N$ expansion and the form-factor bootstrap method \cite{Smirnov}. This approach has been used in Reference \cite{Orland}, to find the form factors of the renormalized field operator.
We also find the two-particle form factor for all $N>2$.

In the next section, we review the exact S matrix for the chiral model. We calculate the two-particle form factors in the planar limit in Section III, and for general $N$ in Section VI. In Section V we calculate the four-particle form factor, and we discuss our results in the final section.
\section{The exact S-matrix and multiparticle states}

The sigma model has elementary particles of mass $m$, which carry both left and right colors. These elementary particles form bound states that obey a sine formula \cite{Schroer}
 \beq
 m_r=m\frac{\sin(\frac{\pi r}{N})}{\sin(\frac{\pi}{N})},\,\,r=1,...,N-1,
 \eeq
where $m_r$ is the mass of a $r$-particle bound state. In the large-$N$ limit, the mass of a r-particle bound state is $m_r=mr$, for finite $r$. This means that there are no bound states of a finite number of elementary particles in the planar limit, since the binding energy vanishes. 
 
We introduce particle and antiparticle creation operators $\mathfrak{A}^\dag_P(\theta)_{ab}$ and 
$\mathfrak{A}^\dag_A(\theta)_{ba}$, respectively, where $\theta$ is the particle rapidity, defined in terms of the momentum vector by $p_0=m\cosh \theta,\,p_1=m\sinh \theta$, and $a,b=1,\dots,N$ are left and right color indices, respectively. A product of creation operators acting on the vacuum in order of increasing rapidity, from left to right, gives the multiparticle state
\beq
|P,\theta_1,a_1,b_1;A,\theta_2,b_2,a_2;...\rangle_{\rm in}=\mathfrak{A}^\dag_P(\theta_1)_{a_1b_2}\mathfrak{A}^\dag_A(\theta_2)_{b_2a_2}\dots|0\rangle,\,\,{\rm where}\,\,\theta_1>\theta_2>\dots\,.
\eeq

The S matrix of two particles, with incoming rapidities $\theta_1$ and $\theta_2$, outgoing rapidities $\theta_1'$ and $\theta_2'$, is
 \beq
 \left._{\rm out}\langle P,\theta_1',c_1,d_1;P,\theta_2',c_2,d_2|P,\theta_1,a_1,b_1;P,\theta_2,a_2,b_2\rangle\right._{\rm in}=S_{PP}(\theta)^{c_2d_2;c_1d_1}_{a_1b_1;a_2b_2}4\pi\delta(\theta_1'-\theta_1)4\pi\delta(\theta_2'-\theta_2),\nonumber
 \eeq
where $\theta=\theta_1-\theta_2$. 
We follow convention and call the function $S_{PP}(\theta)^{c_2d_2;c_1d_1}_{a_1b_1;a_2b_2}$ the S matrix. It is given by
\beq
S_{PP}(\theta)^{c_2d_2;c_1d_1}_{a_1b_1;a_2b_2}=\chi(\theta) S_{\rm CGN}(\theta)^{c_2;c_1}_{a_1;a_2}  S_{\rm CGN}(\theta)^{d_2;d_1}_{b_1;b_2},\label{smatrix}
\eeq
where $S_{\rm CGN}(\theta)$ is the S matrix of two elementary excitations of the $SU(N)$ chiral Gross-Neveu model \cite{Berg},
\cite{Kurak}:
\beq
S_{\rm CGN}(\theta)^{c_2;c_1}_{a_1;a_2}=\frac{\Gamma(i\theta/2\pi+1)\Gamma(-i\theta/2\pi-1/N)}{\Gamma(i\theta/2\pi+1-1/N)\Gamma(-i\theta/2\pi)}\left(\delta^{c_1}_{a_1}\delta^{c_2}_{a_2}-\frac{2\pi i}{N\theta}\delta^{c_1}_{a_2}\delta^{c_2}_{a_1}\right),\nonumber
\eeq
and $\chi(\theta)$ is the CDD factor \cite{Zamolodchikov}:
\beq
\chi(\theta)=\frac{\sinh\left(\frac{\theta}{2}-\frac{\pi i}{N}\right)}{\sinh\left(\frac{\theta}{2}+\frac{\pi i}{N}\right)}.
\eeq
 
The particle-antiparticle S matrix is related to the particle-particle S matrix by crossing, {\em i.e.} 
$\theta\to\hat{\theta}=\pi i-\theta$. The S matrix for a particle with incoming rapidity $\theta_1$ and outgoing rapidity $\theta_1'$,  and an antiparticle with incoming rapidity $\theta_2$  and outgoing rapidity $\theta_2'$ is
\beq
S_{AP}(\theta\!\!\!\!\!&)&\!\!\!\!\!^{d_2 c_2; c_1 d_1}_{a_1b_2;b_2a_2}=S(\hat{\theta},N)\nonumber\\
&\times&\left[\delta^{d_2}_{b_2}\delta^{c_2}_{a_2}\delta^{c_1}_{a_1}\delta^{d_1}_{b_1}-\frac{2\pi i}{N\hat{\theta}}\left(\delta_{a_1a_2}\delta^{c_1c_2}\delta^{d_2}_{b_2}\delta^{d_1}_{b_1}+\delta^{c_2}_{a_2}\delta^{c_1}_{a_1}\delta_{b_1b_2}\delta^{d_1d_2}\right)-\frac{4\pi^2}{N^2\hat{\theta}^2}\delta_{a_1a_2}\delta^{c_1c_2}\delta_{b_1b_2}\delta^{d_1d_2}\right],\label{particleantiparticles}
\eeq
where 
\beq
S(\theta,N)=\frac{\sinh\left(\frac{\theta}{2}-\frac{\pi i}{N}\right)}{\sinh\left(\frac{\hat{\theta}}{2}+\frac{\pi i}{N}\right)}\left[\frac{\Gamma(i\theta/2\pi+1)\Gamma(-i\theta/2\pi-1/N)}{\Gamma(i\theta/2\pi+1-1/N)\Gamma(-i\theta/2\pi)}\right]^2=1+\mathcal{O}\left(\frac{1}{N^2}\right).\label{esfunction}
\eeq

The creation operators satisfy the Zamolodchikov algebra:
\beq
\mathfrak{A}^\dag_P(\theta_1)_{a_1b_1}\mathfrak{A}^\dag_P(\theta_2)_{a_2b_2}&=&S_{PP}(\theta)^{c_2d_2;c_1d_1}_{a_1b_1;a_2b_2}\mathfrak{A}^\dag_P(\theta_2)_{c_2d_2}\mathfrak{A}^\dag_P(\theta_1)_{c_1d_1},\nonumber\\
\mathfrak{A}^\dag_A(\theta_1)_{b_1a_1}\mathfrak{A}^\dag_A(\theta_2)_{b_2a_2}&=&S_{AA}(\theta)^{d_2c_2;d_1c_1}_{b_1a_1;b_2a_2}\mathfrak{A}^\dag_A(\theta_2)_{d_2c_2}\mathfrak{A}^\dag_A(\theta_1)_{d_1c_1},\nonumber\\
\mathfrak{A}^\dag_P(\theta_1)_{a_1b_1}\mathfrak{A}^\dag_A(\theta_2)_{b_2a_2}&=&S_{AP}(\theta)^{d_2c_2;c_1d_1}_{a_1b_1;b_2a_2}\mathfrak{A}^\dag_A(\theta_2)_{d_2c_2}\mathfrak{A}^\dag_P(\theta_1)_{c_1d_1}.\label{zamoalgebra}
\eeq

The $r$-excitation form factor of an operator $\mathfrak{B}(x)$ is defined as
\beq
\langle 0|\mathfrak{B}(x)|I_1,\theta_1,C_1;\dots;I_r,\theta_r,C_r\rangle=e^{-i\sum_{k=1}^r x\cdot p_k}\mathcal{F}^{\mathfrak{B}}_{C_1,\dots C_r}(\theta_1,\dots,\theta_r),\nonumber
\eeq
where $I_k=P$ if the $k^{\rm th}$ excitation is a particle, and 
$I_k=A$ if the $k^{\rm th}$ excitation is an antiparticle,
$C_k$ is the set of indices $a_k,b_k$ for $I_k=P$ or $b_k,a_k$ for $I_k=A$. The $x$-dependence  of the form factor is trivial, due to Lorentz invariance.

The vacuum expectation value of two operators $\mathfrak{B}(x)$ and $\mathfrak{C}(y)$ can be expressed in terms of form factors, using completeness of in states:
\beq
\langle 0|\mathfrak{B}(x)\!\!\!\!&\mathfrak{C}&\!\!\!\!\!(y)|0\rangle=\langle 0|\mathfrak{B}(x)|0\rangle
\langle 0|\mathfrak{C}(y)|0\rangle\nonumber\\
&+&\sum_{r=1}^{\infty}\sum_{t=1}^{\infty}\int\frac{d\theta_1\dots d\theta_r,d\phi_1\dots d\phi_t}{(2\pi)^
{r+t}(r+t)!}\langle 0|\mathfrak{B}(x)|P,\theta_1,a_1,b_1;\dots;P,\theta_r,a_r,b_r;A,\phi_1,d_1,c_1;\dots;A,\phi_t,d_t,c_t\rangle\nonumber\\
&\times&\langle P,\theta_1,a_1,b_1;\dots;P,\theta_r,a_r,b_r;A,\phi_1,d_1,c_1;\dots;A,\phi_t,d_t,c_t|\mathfrak{C}(y)|0\rangle.\label{wightman}
\eeq

\section{Smirnov's axioms and the two-particle form factors}

In this section we calculate the first nonvanishing form factor of the current operators at large $N$. 
We will discuss only the left-handed current $j^L_\mu(x)^c_a$ in detail, since the same method yields the right-handed-current form factor.

Under a global $SU(N)\times SU(N)$ transformation, the current and the particle and antiparticle creation operators transform as
\beq
j^L_\mu(x)\to V_L j^L_\mu(x) V^\dag_L,\,\,\,\mathfrak{A}^\dag_P(\theta)\to V_R^\dag\mathfrak{A}^\dag_P(\theta)V^\dag_L,\,\,\,\mathfrak{A}^\dag_A(\theta)\to V_L\mathfrak{A}^\dag_A(\theta)V_R.\nonumber
\eeq
Only form factors with equal number of particles and antiparticles are invariant under such global transformations. The first nontrivial form factor 
is
\beq
\langle 0|j^L_\mu(x)_{a_0c_0}\!\!\!\!\!&|&\!\!\!\!\!A,\theta_1,b_1,a_1;P,\theta_2,a_2,b_2\rangle=\langle 0| j^L_\mu(x)_{a_0c_0}\mathfrak{A}^\dag_{A}(\theta_1)_{b_1a_1}\mathfrak{A}^\dag_P(\theta_2)_{a_2b_2}|0\rangle\nonumber\\
&=&(p_1-p_2)_\mu e^{-ix\cdot (p_1+p_2)}\left[F_1(\theta)\delta_{a_0a_2}\delta_{b_1b_2}\delta_{c_0a_1}+F_2(\theta)\delta_{a_0c_0}\delta_{b_1b_2}\delta_{a_1a_2}\right],\label{ffunctions}
\eeq
for $\theta_1>\theta_2$, and
\beq
\langle 0|j^L_\mu(x)_{a_0c_0}\!\!\!\!\!&|&\!\!\!\!\!P_1,\theta_1,a_1,b_1;A,\theta_2,b_2,a_2\rangle=\langle 0| j^L_\mu(x)_{a_0c_0}\mathfrak{A}^\dag_{P}(\theta_2)_{a_2b_2}\mathfrak{A}^\dag_A(\theta_1)_{b_1a_1}|0\rangle\nonumber\\
&=&(p_1-p_2)_\mu e^{-ix\cdot (p_1+p_2)}\left[F_1'(\theta)\delta_{a_0a_2}\delta_{b_1b_2}\delta_{c_0a_1}+F_2'(\theta)\delta_{a_0c_0}\delta_{b_1b_2}\delta_{a_1a_2}\right],
\eeq
for $\theta_2>\theta_1$, where, as before, $\theta=\theta_1-\theta_2$. Lorentz invariance requires that the functions $F_1(\theta)$ and $F_2(\theta)$ depend only on the rapidity difference $\theta$
\cite{Smirnov}. 

We next apply the scattering axiom, also known as Watson's theorem \cite{Smirnov}. This axiom follows from the Zamolodchikov algebra (\ref{zamoalgebra}) on the creation operators of the in-state. This gives a relation between $F_{1,2}(\theta)$ and $F_{1,2}'(\theta)$:
\beq
\langle 0|j^L_\mu(0)_{a_0c_0}\mathfrak{A}^\dag_P(\theta_2)_{a_2b_2}\mathfrak{A}^\dag_A(\theta_1)_{b_1a_1}|0\rangle=S_{AP}(\theta)^{d_1c_1;c_2d_2}_{a_2b_2;b_1a_1}\langle 0| j^L_\mu(0)_{a_0c_0}\mathfrak{A}^\dag_{A}(\theta_1)_{d_1c_1}\mathfrak{A}^\dag_P(\theta_2)_{c_2d_2}|0\rangle.\label{Wat}
\eeq
After some work, these reduce to the relations
\beq
F_1'(\theta)&=&S(\hat{\theta},N)\left(1-\frac{2\pi i}{\hat{\theta}}\right)F_1(\theta),\nonumber\\
F_2'(\theta)&=&S(\hat{\theta},N)\left[\left(1-\frac{2\pi i}{\hat{\theta}}\right)^2F_2(\theta)+\frac{1}{N}\left(\frac{-2\pi i}{\hat{\theta}}-\frac{4\pi^2}{\hat{\theta}^2}\right)F_1(\theta)\right]. \label{watson}
\eeq
In obtaining (\ref{watson}), some factors of $1/N$ in the S matrix were canceled by summing over group indices in (\ref{Wat}).

We next consider the Smirnov periodicity axiom \cite{Smirnov}, which follows from crossing symmetry. For the $M$-excitation form factor of an operator $\mathfrak{B}(0)$, the periodicity axiom is
\beq
\langle 0|\mathfrak{B}(0)\mathfrak{A}^\dag_{I_1}\!\!\!\!\!&(&\!\!\!\!\!\theta_1)_{C_1}\mathfrak{A}^\dag_{I_1}(\theta_2)_{C_2}\dots \mathfrak{A}^\dag_{I_M}(\theta_M)_{C_M}|0\rangle\nonumber\\
&=&\langle 0|\mathfrak{B}(0)\mathfrak{A}^\dag_{I_M}(\theta_M-2\pi i)_{C_M}\mathfrak{A}^\dag_{I_1}(\theta_1)_{C_1}\dots \mathfrak{A}^\dag_{I_{M-1}}(\theta_{M-1})_{C_{M-1}}|0\rangle. \label{periodicity}
\eeq
For more discussion of this axiom see References \cite{Orland}, \cite{Babujian}. 

Applying the periodicity axiom to our form factors (\ref{ffunctions}), we find the two equivalent conditions:
\beq
\langle 0| j_\mu^L(0)_{a_0c_0}\mathfrak{A}^\dag_P(\theta_2)_{a_2b_2}\mathfrak{A}^\dag_A(\theta_1)_{b_1a_1}| 0 \rangle&=&\langle 0|j^L_\mu(0)_{a_0c_0}\mathfrak{A}^\dag_{A}(\theta_1-2\pi i)_{b_1c_1}\mathfrak{A}^\dag_P(\theta_2)_{a_2b_2}|0\rangle \nonumber\\
\Rightarrow F_{1,2}'(\theta)&=&F_{1,2}(\theta-2\pi i),\label{periodicityone}
\eeq
and
\beq
\langle 0|j^L_\mu(0)_{a_0c_0}\mathfrak{A}^\dag_A(\theta_1)_{b_1a_1}\mathfrak{A}^\dag_P(\theta_2)_{a_2b_2}|0\rangle&=&\langle 0|j^L_\mu(0)_{a_0c_0}\mathfrak{A}^\dag_P(\theta_2-2\pi i)_{a_2b_2}\mathfrak{A}^\dag_A(\theta_1)_{b_1a_1}|0\rangle\nonumber\\
\Rightarrow F_{1,2}(\theta)&=&F_{1,2}'(\theta+2\pi i).\label{periodicitytwo}
\eeq 
Combining (\ref{watson}) with (\ref{periodicityone}) gives 
\beq
F_1(\theta-2\pi i)&=&\hat{S}(\theta,N)\left(\frac{\theta+\pi i}{\theta-\pi i}\right)F_1(\theta),\nonumber\\
F_2(\theta-2\pi i)&=&\hat{S}(\theta,N)\left(\frac{\theta+\pi i}{\theta-\pi i}\right)^2F_2(\theta)+\hat{S}(\theta,N)\frac{2\pi i}{N(\theta-\pi i)}\left(\frac{\theta+\pi i}{\theta-\pi i}\right)F_1(\theta),\label{functional}
\eeq
where we have defined the function $\hat{S}(\theta,N)\equiv S(\hat{\theta},N)$.

The tracelessness of the current operator implies
\beq
\langle 0|j^L_\mu(x)_a^a|A,\theta_1,b_1,a_1;P,\theta_2,a_2,b_2\rangle=(p_1-p_2)_\mu\left[F_1(\theta)\delta_{a_1a_2}\delta_{b_1b_2}+NF_2(\theta)\delta_{a_1a_2}\delta_{b_1b_2}\right]=0,\nonumber
\eeq
or
\beq
F_2(\theta)=-\frac{F_1(\theta)}{N}.\label{traceless}
\eeq 

The form factor is now
\beq
\langle 0|j^L_\mu(0)_{a_0c_0}\!\!\!\!\!&|&\!\!\!\!\!A,\theta_1, b_1,a_1;P,\theta_2,a_2,b_2\rangle\nonumber\\
&=&\left(p_1-p_2\right)_\mu F_1(\theta)\left[\delta_{a_0a_2}\delta_{b_1b_2}\delta_{c_0a_1}-\frac{1}{N}\delta_{a_0c_0}\delta_{b_1b_2}\delta_{a_1a_2}\right],\label{tracelessformfactor}
\eeq
where $F_1(\theta)$ satisfies
\beq
F_1(\theta-2\pi i)=\hat{S}(\theta,N)\left(\frac{\theta+\pi i}{\theta-\pi i}\right)F_1(\theta).\label{efone}
\eeq

For large $N$, we expand $\hat{S}(\theta,N)=1+\mathcal{O}\left(\frac{1}{N^2}\right)$ and $F_1(\theta)=F_1^0(\theta)+\frac{1}{N}F_1^1(\theta)+\frac{1}{N^2}F_1^2(\theta)+\dots$, so that
\beq
F_1^0(\theta-2\pi i)=\left(\frac{\theta+\pi i}{\theta-\pi i}\right)F_1^0(\theta).\label{planar}
\eeq
The general solution to (\ref{planar}) is 
\beq
F_1^0(\theta)=\frac{g(\theta)}{\theta+\pi i},\label{solutionplanar}
\eeq
where $g(\theta)$ satisfies the periodicity condition $g(\theta-2\pi i)=g(\theta)$.
The minimal choice is to take $g(\theta)=g$, a constant.

Next we determine the value of $g$. There is a conserved charge $Q^L_{a_0c_0}$, associated with the current operator. This charge is 
\beq
Q^L_{a_0c_0}=\int dx^1 j^L_0(x)_{a_0c_0}.\nonumber
\eeq
We fix the value of $g$ by 
requiring that the charge generates the $SU(N)$ Lie algebra:
\beq
Q^{L\,a}_a=0,\,\,\,\,\left[Q^{L\,c_1}_{a_2},Q^{L\,c_2}_{a_2}\right]=if^{c_1c_2a_3}_{a_1a_2c_3}Q^{L\,c_3}_{a_3},\label{sunalgebra}
\eeq
where the structure coefficients are
\beq
f^{c_1c_2a_3}_{a_1a_2c_3}=i\left(\delta^{c_2}_{a_1}\delta^{a_3}_{a_2}\delta^{c_1}_{c_3}-\delta^{c_1}_{a_2}\delta^{a_3}_{a_1}\delta^{c_2}_{c_3}\right).\nonumber
\eeq
We cross the incoming particle from Equation (\ref{tracelessformfactor}) to an outgoing antiparticle, 
{\em via} $\theta_2\to \theta_2-\pi i$, to find
\beq
\,\langle A,\theta_2,b_2,a_2\!\!\!\!\!&|&\!\!\!\!\! j^L_0(x)_{a_0c_0}|A, \theta_1,b_1,a_1\rangle\nonumber\\
&=&m(\cosh \theta_1+\cosh \theta_2)\exp \{-im[x^0(\cosh \theta_1-\cosh \theta_2)-x^1(\sinh \theta_1-\sinh \theta_2)]\}\nonumber\\
&&\times F_1(\theta+\pi i)\left(\delta_{a_0a_2}\delta_{b_1b_2}\delta_{c_0a_1}-\frac{1}{N}\delta_{a_0c_0}\delta_{b_1b_2}\delta_{a_1a_2}\right).\nonumber
\eeq
The integral over $x^1$ gives the matrix element of the charge operator:
\beq
\langle A,\theta_1,b_2,a_2\!\!\!\!\!&|&\!\!\!\!\! Q^L_{a_0c_0}|A,\theta_1,b_1,a_1\rangle\nonumber\\
&=&(2\pi)^22(p_1)_0\delta(\theta_1-\theta_2)\left(\delta_{a_0a_2}\delta_{b_1b_2}\delta_{c_0a_1}-\frac{1}{N}\delta_{a_0c_0}\delta_{b_1b_2}\delta_{a_1a_2}\right)F_1(\pi i).\nonumber
\eeq
The matrix element of the commutator of two charges is found by inserting a complete set of one-antiparticle intermediate states:
\beq
\langle A,\theta_2,b_2,a_2\!\!\!\!\!&|&\!\!\!\!\! \left[Q^L_{a_0c_0},Q^L_{a_4c_4}\right]|A,\theta_1,b_1,a_1\rangle\nonumber\\
&=&\int\frac{d\theta_3}{4\pi}\langle A,\theta_2,b_2,a_2|Q^L_{a_0c_0}|A,\theta_3,b_3,a_3\rangle\langle A,\theta_3,b_3,a_3|Q^L_{a_4c_4}|A,\theta_1,b_1,a_1\rangle\nonumber\\
&&-\int\frac{d\theta_3}{4\pi}\langle A,\theta_2,b_2,a_2|Q^L_{a_4c_4}|A,\theta_3,b_3,a_3\rangle\langle A,\theta_3,b_3,a_3|Q^L_{a_0c_0}|A,\theta_1,b_1,a_1\rangle.\label{commutator}
\eeq
With the 
choice $F(\pi i)=1$, Equation (\ref{commutator}) becomes
\beq
\langle A,\theta_2,b_2,a_2\!\!\!\!\!&|&\!\!\!\!\! \left[Q^{L\,c_0}_{a_0},Q^{L\,c_4}_{a_4}\right]|A,\theta_1,b_1,a_1\rangle\nonumber\\
&=&if^{c_0c_4a_5}_{a_0a_4c_5}\langle A,\theta_2,b_2,a_2|Q^{L\, c_5}_{a_5}|A,\theta_1,b_1,a_1\rangle,\nonumber
\eeq
which is equivalent to (\ref{sunalgebra}). This 
fixes the constant $g=2\pi i$.

We have not yet discussed the annihilation-pole axiom \cite{Smirnov}. 
This axiom relates the form factors of $M$ particles to the form factors of $M-2$ particles. 
The general multiparticle form factor of the current operator is
\beq
\langle 0 \!\!\!\!\!&| &\!\!\!\!\!  j_\mu^L(0)_{a_0c_0}| A, \theta_1, b_1, a_1;\dots ; A,\theta_l,b_l,a_l;P,\theta_{l+1},a_{l+1},b_{l+1};\dots ;P,\theta_{2l},a_{2l},b_{2l};A,\theta_{n-1},b_{n-1},a_{n-1};P,\theta_{n},a_n,b_n\rangle \nonumber\\
&=&\left[p_1+\dots+p_l-(p_{l+1}+\dots + p_{2l})+p_{n-1}-p_{n}\right]_\mu \mathcal{F}^{\mathcal{O}}(\theta_1,\dots,\theta_n)_{a_0c_0a_1\dots a_n;b_1\dots b_n}.\label{generalformfactor}
\eeq
Here we have factored out the vector-valued pre-factor in square brackets, consisting of a linear combination of the particle momenta, chosen to
make $ \mathcal{F}^{\mathcal{O}}(\theta_1,\dots,\theta_n)$ a Lorentz scalar.
We 
define a Lorentz-scalar-valued operator $\mathcal{O}_{a_0c_0}$ by
\beq
&\langle 0| \mathcal{O}_{a_0c_0}| A,\theta_1,b_1,a_1;\dots;A,\theta_l,b_l,a_l;P,\theta_{l+1},a_{l+1},b_{l+1};\dots;P,\theta_{2l},a_{2l},b_{2l};A,\theta_{n-1},b_{n-1},a_{n-1};P,\theta_n,a_n,b_n\rangle \nonumber\\
&\equiv\mathcal{F}^{\mathcal{O}}(\theta_1,\dots,\theta_n)_{a_0c_0a_1\dots a_n;b_1\dots b_n}. \nonumber
\eeq
The form factor has a pole at $\theta_{n-1\,n}\equiv\theta_{n-1}-\theta_{n}=-\pi i$, corresponding to 
annihilation of the $(n-1)^{\rm st}$ and $n^{\rm th}$ excitations. We cross the $n^{\rm th}$ particle to an outgoing antiparticle, yielding
\beq
\langle A, \theta_n\!\!\!\!\!&,&\!\!\!\!\!  b_n, a_n| j_\mu^L(0)_{a_0c_0}| A, \theta_1,b_1,a_1;\dots;A,\theta_l,b_l,a_l;P, \theta_{l+1},a_{l+1},b_{l+1};\dots;P,\theta_{2l},a_{2l},b_{2l};A,\theta_{n-1},b_{n-1},a_{n-1}\rangle\nonumber\\ 
&=&\left[p_1+\dots+p_l-(p_{l+1}+\dots+p_{2l})+p_{n-1}+p_n\right]_\mu\nonumber\\
&&\times\langle A,\theta_n,b_n,a_n\left| \mathcal{O}_{a_0c_0}\right|A,\theta_1,b_1,a_1;\dots;A,\theta_l,b_l,a_l;P,\theta_{l+1},a_{l+1},b_{l+1};\dots;P,\theta_{2l},a_{2l},b_{2l};A,\theta_{n-1},b_{n-1},a_{n-1}\rangle.\nonumber
\eeq
By the generalized crossing formula \cite{Babujian},
\beq
\langle A,\theta_n,b_n,a_n\!\!\!\!\!&|&\!\!\!\!\!  \mathcal{O}_{a_0c_0}|A,\theta_1,b_1,a_1;\dots;A,\theta_l,b_l,a_l;P,\theta_{l+1},a_{l+1},b_{l+1};\dots;P,\theta_{2l},a_{2l},b_{2l};A,\theta_{n-1},b_{n-1},a_{n-1}\rangle\,\,\,\,\,\,\,\,\,\,\,\,\,\,\nonumber\\
\nonumber\\
&=&\langle A,\theta_n,b_n,a_n|A,\theta_1,b_1,a_1\rangle\mathcal{F}^{\mathcal{O}}(\theta_2,\dots,\theta_{n-1})_{a_0c_0a_2\dots a_{n-1};b_2\dots b_{n-1}}\nonumber\\
&&+\mathcal{F}^{\mathcal{O}}(\theta_n-i\pi_{-},\theta_1,\dots,\theta_{n-1})_{a_0c_0a_na_1\dots,a_{n-1};b_nb_1\dots b_{n-1}}\,\,\,\,\,\,{\rm for}\,\,\,\theta_n\geq\theta_1>\dots>\theta_{n-1}\nonumber\\
{\rm or} \,\,\,\,\,\,\,\,\,\,\,\,\,\,\,\,\,\,\,\,\,\,\,\,\,\,\,\,\,\,\,\,\,\,\,\,\,\,\,\,\,\,\,\,\,&&\nonumber\\
 &=&\langle A,\theta_n,b_n,a_n|A,\theta_{n-1},b_{n-1},a_{n-1}\rangle \mathcal{F}^{\mathcal{O}}(\theta_1,\dots,\theta_{2l})_{a_0c_0a_1\dots a_{2l};b_1\dots b_{2l}}\nonumber\\
&& +\mathcal{F}^{\mathcal{O}}(\theta_1,\dots,\theta_{n-1},\theta_{n}+i\pi_{-})_{a_0c_0a_1\dots a_n;b_1\dots b_n}\,\,\,\,\,\,\,\,\,\,\,\,\,\,\,\,\,\,\,\,\,\,\,\,\,\,\,\,\,{\rm for}\,\,\,\theta_1>\dots>\theta_{n-1}\geq \theta_n,\label{generalcrossing}
\eeq
where the right-hand side contains the $n$- and the $n-2$-particle form factors, and $\pi_{-}=\pi-\epsilon$.
Near the annihilation pole at $\theta_{n-1\,n}=-\pi i$ the form factors are of the form:
\beq
\mathcal{F}^{\mathcal{O}}(\theta_{n}-i\pi_{-},\theta_1,\dots,\theta_{n-1})_{a_0c_0a_1\dots a_n;b_1\dots b_n}&=&\frac{1}{\theta_{n-1}-\theta_n+i\epsilon}\,h(\theta_1,\dots,\theta_n)_{a_0c_0a_1\dots a_n;b_1\dots b_n},\nonumber\\
&{\rm and}&\nonumber\\
\mathcal{F}^{\mathcal{O}}(\theta_1,\dots,\theta_{n-1},\theta_n+i\pi_{-})_{a_0c_0a_1\dots a_n;b_1\dots b_n}&=&\frac{1}{\theta_{n-1}-\theta_n-i\epsilon}\, h(\theta_1,\dots,\theta_n)_{a_0c_0a_1\dots a_n;b_1\dots b_n},\nonumber
\eeq
where $h(\theta_1,\dots,\theta_n)_{a_0c_0a_1\dots a_n;b_1\dots b_n}$ is an analytic function in $\theta_{n-1\,n}$. We use the identity
\beq
\frac{1}{\theta_{n-1}-\theta_n\pm i\epsilon}=\mathbf{P}\left\{\frac{1}{\theta_{n-1}-\theta_n}\right\}\mp i\pi\delta(\theta_{n-1}-\theta_n),\nonumber
\eeq
where $\mathbf{P}\left\{f(\theta_{n-1},\theta_n)\right\}$ is the principal value of $f(\theta_{n-1},\theta_n)$.
We apply Watson's theorem to Equation (\ref{generalcrossing}), and find
\beq
\langle A,\theta_n,b_n,a_n\!\!\!\!\!&|&\!\!\!\!\!  \mathcal{O}_{a_0c_0}|A,\theta_1,b_1,a_1;\dots;A,\theta_l,b_l,a_l;P,\theta_{l+1},a_{l+1},b_{l+1};\dots;P,\theta_{2l},a_{2l},b_{2l};A,\theta_{n-1},b_{n-1},a_{n-1}\rangle\nonumber\\
&=&\langle A,\theta_n,b_n,a_n|A,\theta_{n-1},b_{n-1}',a_{n-1}'\rangle S_{AA}(\theta_{1\,n-1})^{b_{n-1}'a_{n-1}';b_1'a_1'}_{d_1c_1;b_1a_1}\times\dots\times S_{AA}(\theta_{l\,n-1})^{d_{l-1}c_{l-1};b_l'a_l'}_{d_lc_l;b_lc_l}\nonumber\\
&&\times S_{AP}(\theta_{n-1\,l+1})^{d_lc_l;a_{l+1}'b_{l+1}'}_{c_{l+1}d_{l+1};a_{l+1}b_{l+1}}\times\dots\times S_{AP}(\theta_{n-1\,2l})^{c_{2l-1}d_{2l-1};a_{2l}'b_{2l}'}_{c_{2l}d_{2l};a_{2l}b_{2l}}\times\mathcal{F}^{\mathcal{O}}(\theta_1,\dots,\theta_{2l})_{a_0c_0a_1'\dots a_{2l}';b_1'\dots b_{2l}'}\nonumber\\
&&+\left(\mathbf{P}\left\{\frac{1}{\theta_{n-1}-\theta_n}\right\}-i\pi \delta(\theta_{n-1}-\theta_n)\right)\,h(\theta_1,\dots,\theta_n)_{a_0c_0a_1\dots a_n;b_1\dots b_n}\nonumber\\
&=&\langle A,\theta_n,b_n,a_n|A,\theta_{n-1},b_{n-1},a_{n-1}\rangle \mathcal{F}^{\mathcal{O}}(\theta_1,\dots,\theta_{2l})_{a_0c_0a_1\dots a_{2l};b_1\dots b_{2l}}\nonumber\\
&&+\left(\mathbf{P}\left\{\frac{1}{\theta_{n-1}-\theta_n}\right\}+i\pi\delta(\theta_{n-1}-\theta_n)\right)\,h(\theta_1,\dots,\theta_n)_{a_0c_0a_1\dots a_n;b_1\dots b_n}.\label{watsoncrossing}
\eeq
We will use the normalization $\langle A,\theta_n,b_n,a_n|A,\theta_{n-1},b_{n-1},a_{n-1}\rangle=4\pi\delta_{a_{n-1}a_n}\delta_{b_{n-1}b_n}\delta(\theta_{n-1}-\theta_n).$ Comparing the terms proportional to $\delta(\theta_{n-1}-\theta_n)$ in (\ref{watsoncrossing}), we recover the annihilation pole axiom \cite{Babujian}:
\beq
h(\theta_1,\dots, \!\!\!&\theta_{n-1}&\!\!\!,\theta_{n-1})_{a_0c_0a_1\dots a_n;b_1\dots b_n}\nonumber\\
&=&{\rm Res}|_{\theta_{n-1\,n}=-\pi i}\mathcal{F}^{\mathcal{O}}(\theta_1,\dots,\theta_{2l},\theta_{n-1},\theta_n)_{a_0c_0a_1\dots a_{2l}a_{n-1}a_n;b_1\dots b_{2l}b_{n-1}b_n}\nonumber\\
&=&2i\,\mathcal{F}^{\mathcal{O}}(\theta_1,\dots,\theta_{2l})_{a_0c_0a_1'\dots a_{2l}';b_1'\dots b_{2l}'}\delta_{a_{n-1}'a_n}\delta_{b_{n-1}'b_n}\nonumber\\
&&\times\left(\delta_{a_1'a_1}\dots\delta_{a_{n-1}'a_{n-1}}\delta_{b_1'b_1}\dots\delta_{b_{n-1}'b_{n-1}}-S_{AA}(\theta_{1\,n-1})^{b_{n-1}'a_{n-1}';b_1'a_1'}_{d_1c_1;b_1a_1}\times\dots\times S_{AA}(\theta_{l\,n-1})^{d_{l-1}c_{l-1};b_l'a_l'}_{d_lc_l;b_lc_l}\right.\nonumber\\
&&\times \left.S_{AP}(\theta_{n-1\,l+1})^{d_lc_l;a_{l+1}'b_{l+1}'}_{c_{l+1}d_{l+1};a_{l+1}b_{l+1}}\times\dots\times S_{AP}(\theta_{n-1\,2l})^{c_{2l-1}d_{2l-1};a_{2l}'b_{2l}'}_{c_{2l}d_{2l};a_{2l}b_{2l}}\right).\label{annihilationpoleaxiom}
\eeq

\section{Two-particle form factors at finite $N$}

In this section, we find the exact two-particle form factor of the current operator, for arbitrary $N\geq 2$. For $N=2$, the principal chiral model is equivalent to an ${O}(4)$-symmetric vector model. The form factors of currents of the ${O}(4)$ model were found in Reference \cite{KarowskiandWeisz}. 
 
Our result for the two-particle form factor, for general $N$, is
\beq
\langle 0|j^L_\mu(0)_{a_0c_0}\!\!\!\!\!&|&\!\!\!\!\! A,\theta_1,b_1,a_1;P,\theta_2,a_2,b_2\rangle\nonumber\\
&=&(p_1-p_2)_\mu F_1(\theta)\left(\delta_{a_0a_2}\delta_{b_1b_2}\delta_{c_0a_1}-\frac{1}{N}\delta_{a_0c_0}\delta_{b_1b_2}\delta_{a_1a_2}\right)\nonumber,
\eeq
where $F_1(\theta)$ satisfies equation (\ref{efone}).
We insert
\beq
F_1(\theta)=\frac{g(\theta)}{\theta+\pi i},\nonumber
\eeq
into (\ref{efone}), finding
\beq
g(\theta-2\pi i)=\hat{S}(\theta,N)g(\theta).\label{equationg}
\eeq
We solve Equation (\ref{equationg}) by a contour-integration method first used in Reference \cite{KarowskiandWeisz}. We define a contour $C$ to be that from $-\infty$ to $\infty$ and from $\infty+2\pi i$ to $-\infty+2\pi i$, bounding the strip in which the form factor is holomorphic. Then
\beq
\ln g(\theta)=\int_C\frac{dz}{4\pi i}\coth \frac{z-\theta}{2}\ln g(z)=\int_{-\infty}^\infty \frac{dz}{4\pi i}\coth \frac{z-\theta}{2}\ln\frac{g(z)}{g(z+2\pi i)}.\nonumber
\eeq
We differentiate both sides with respect to $\theta$, and use (\ref{equationg}) to write
\beq
\frac{d}{d\theta}\left[\ln g(\theta)\right]=\frac{1}{8\pi i}\int_{-\infty}^\infty\frac{dz}{\sinh^2\frac{1}{2}(z-\theta)}\ln \hat{S}(z,N).\label{contourintegral}
\eeq
 The solution to (\ref{contourintegral}) is
\beq
g(\theta)=g\exp \int_0^\infty dx \,A(x,N)\frac{\sin^2[x(\pi i-\theta)/2\pi]}{\sinh x},\label{minimalform}
\eeq
where the function $A(x,N)$ is defined by
\beq
\hat{S}(\theta,N)=\exp \int_0^\infty dx \,A(x,N) \sinh \left(\frac{x\theta}{\pi i}\right),\label{exponentialexpression}
\eeq
and $g$ is a constant. Note that expanding the $S$ matrix in powers of $1/N$ yields $A(x,N)=\frac{1}{N^2}B(x)+\mathcal{O}(\frac{1}{N^3})$.

To express the function $\hat{S}(\theta,N)$, presented in (\ref{esfunction}), in the form 
(\ref{exponentialexpression}), we use the integral formula of the gamma function 
\cite{WhittakerandWatson}, \cite{Weisz},
\beq
\Gamma(z)=\exp\int_0^\infty\frac{dx}{x}\left[\frac{e^{-x z}-e^{-x}}{1-e^{-x}}+(z-1)e^{-x}\right],\,\,\,{\rm for}\,{\rm Re}\,z>0.\nonumber
\eeq
Then
\beq
\left[\frac{\Gamma\left(\frac{i\hat{\theta}}{2\pi}+1\right)\Gamma\left(\frac{-i\hat{\theta}}{2\pi}-\frac{1}{N}\right)}{\Gamma\left(\frac{i\hat{\theta}}{2\pi}+1-\frac{1}{N}\right)\Gamma\left(\frac{-i\hat{\theta}}{2\pi}\right)}\right]^2=\exp\int_0^\infty\frac{dx}{x}\frac{4e^{-x}\left(e^{2x/N}-1\right)}{1-e^{-2x}}\sinh\left(\frac{x\theta}{\pi i}\right),\label{gamma}
\eeq
for $N>2$. We use the formula \cite{Babujian}
\beq
\frac{\sin\frac{\pi}{2}(z+a)}{\sin\frac{\pi}{2}(z-a)}=\exp 2\int_0^\infty\frac{dx}{x}\frac{\sinh x(1-z)}{\sinh x}\sinh(xa),\,\,\,{\rm for}\,\,0<z<1,\nonumber
\eeq
to write the CDD factor as
\beq
\frac{\sinh\left(\frac{\hat{\theta}}{2}-\frac{\pi i}{N}\right)}{\sinh\left(\frac{\hat{\theta}}{2}+\frac{\pi i}{N}\right)}=\frac{\sin\frac{\pi}{2}\left(\left(1-\frac{2}{N}\right)-\frac{\theta}{\pi i}\right)}{\sin\frac{\pi}{2}\left(\left(1-\frac{2}{N}\right)+\frac{\theta}{\pi i}\right)}=\exp\int_0^\infty\frac{dx}{x}\frac{-2\sinh(2x/N)}{\sinh x}\sinh \left(\frac{x\theta}{\pi i}\right),\label{sines}
\eeq
for $N>2$. Combining (\ref{gamma}) and (\ref{sines}) gives
\beq
\hat{S}(\theta,N)=\exp\int_0^\infty \frac{dx}{x}\left[\frac{-2\sinh(2x/N)}{\sinh x}+\frac{4e^{-x}\left(e^{2x/N}-1\right)}{1-2^{-2x}}\right]\sinh\left(\frac{x\theta}{\pi i}\right).\label{afunction}
\eeq
From (\ref{equationg}) and (\ref{minimalform}), the form factor is
\beq
F_1(\theta)=\frac{g}{(\theta+\pi i)}\exp\int_0^\infty \frac{dx}{x}\left[\frac{-2\sinh\left(\frac{2x}{N}\right)}{\sinh x}+\frac{4e^{-x}\left(e^{2x/N}-1\right)}{1-e^{-2x}}\right]\frac{\sin^2[x(\pi i-\theta)/2\pi]}{\sinh x}.\label{finitenform}
\eeq
The condition $F_1(\pi i)=1$ implies $g=2\pi i$.

\section{Four-particle form factors}

Next we find the four-excitation form factor of the current operator, in the large-N limit. Only the form factor with two particles and two antiparticles is non-zero, because of the global symmetry. The most general Lorentz- and $SU(N)\times SU(N)$-invariant  four-particle form factor, respecting the tracelessness of the current operator is
\beq
\langle 0|j_\mu^L(0)_{a_0c_0}\!\!\!\!\!&|&\!\!\!\!\! A,\theta_1,b_1,a_1;A,\theta_2,b_2,a_2;P,\theta_3,a_3,b_3;P,\theta_4,a_4,b_4\rangle\nonumber\\
&=&\langle 0| j_\mu^L(0)_{a_0c_0} \mathfrak{A}_{A}^\dag(\theta_1)_{b_1a_1}\mathfrak{A}_A^\dag(\theta_2)_{b_2a_2}\mathfrak{A}_P^\dag(\theta_3)_{a_3b_3}\mathfrak{A}_P^\dag(\theta_4)_{a_4b_4}|0\rangle\nonumber\\
&=&\frac{1}{N}[p_1+p_2-p_3-p_4]_\mu\,\vec{F}(\theta_1,\theta_2,\theta_3,\theta_4)\cdot \vec{D}_{a_0c_0a_1a_2a_3a_4;b_1b_2b_3b_4},\label{fourparticleone}
\eeq
for $\theta_1>\theta_2>\theta_3>\theta_4$,
\beq
\langle 0|j_\mu^L(0)_{a_0c_0}\!\!\!\!\!&|&\!\!\!\!\! A,\theta_1,b_1,a_1;P,\theta_2,a_2,b_2;A,\theta_3,b_3,a_3;P,\theta_4,a_4,b_4\rangle\nonumber\\
&=&\langle 0|j_\mu^L(0)_{a_0c_0}\mathfrak{A}_{A}^\dag(\theta_1)_{b_1a_1}\mathfrak{A}_P^\dag(\theta_3)_{a_3b_3}\mathfrak{A}_A^\dag(\theta_2)_{b_2a_2}\mathfrak{A}_P^\dag(\theta_4)_{a_4b_4}|0\rangle\nonumber\\
&=&\frac{1}{N}[p_1+p_2-p_3-p_4]_\mu\,\vec{G}(\theta_1,\theta_2,\theta_3,\theta_4)\cdot \vec{D}_{a_0c_0a_1a_2a_3a_4;b_1b_2b_3b_4},\label{fourparticletwo}
\eeq
for $\theta_1>\theta_3>\theta_2>\theta_4$,
\beq
\langle 0|j_\mu^L(0)_{a_0c_0}\!\!\!\!\!&|&\!\!\!\!\! A,\theta_1,b_1,a_1;P,\theta_2,a_2,b_2;P,\theta_3,a_3,b_3;A,\theta_4,b_4,a_4\rangle\nonumber\\
&=&\langle 0|j_\mu^L(0)_{a_0c_0}\mathfrak{A}_{A}^\dag(\theta_1)_{b_1a_1}\mathfrak{A}_P^\dag(\theta_3)_{a_3b_3}\mathfrak{A}_P^\dag(\theta_4)_{a_4b_4}\mathfrak{A}_A^\dag(\theta_2)_{b_2a_2}|0\rangle\nonumber\\
&=&\frac{1}{N}[p_1+p_2-p_3-p_4]_\mu\,\vec{H}(\theta_1,\theta_2,\theta_3,\theta_4)\cdot \vec{D}_{a_0c_0a_1a_2a_3a_4;b_1b_2b_3b_4},\label{fourparticlethree}
\eeq
for $\theta_1>\theta_3>\theta_4>\theta_2$,
\beq
\langle 0|j_\mu^L(0)_{a_0c_0}\!\!\!\!\!&|&\!\!\!\!\! P,\theta_1,a_1,b_1;A,\theta_2,b_2,a_2;P,\theta_3,a_3,b_3;A,\theta_4,b_4,a_4\rangle\nonumber\\
&=&\langle 0|j_\mu^L(0)_{a_0c_0}\mathfrak{A}_{P}^\dag(\theta_3)_{a_3b_3}\mathfrak{A}_A^\dag(\theta_1)_{b_1a_1}\mathfrak{A}_P^\dag(\theta_4)_{a_4b_4}\mathfrak{A}_A^\dag(\theta_2)_{b_2a_2}|0\rangle\nonumber\\
&=&\frac{1}{N}[p_1+p_2-p_3-p_4]_\mu\,\vec{K}(\theta_1,\theta_2,\theta_3,\theta_4)\cdot \vec{D}_{a_0c_0a_1a_2a_3a_4;b_1b_2b_3b_4},\label{fourparticlefour}
\eeq
for $\theta_3>\theta_1>\theta_4>\theta_2$,
\beq
\langle 0|j_\mu^L(0)_{a_0c_0}\!\!\!\!\!&|&\!\!\!\!\! P,\theta_1,a_1,b_1;P,\theta_2,a_2,b_2;A,\theta_3,b_3,a_3;A,\theta_4,b_4,a_4\rangle\nonumber\\
&=&\langle 0|j_\mu^L(0)_{a_0c_0}\mathfrak{A}_{P}^\dag(\theta_3)_{a_3b_3}\mathfrak{A}_P^\dag(\theta_4)_{a_4b_4}\mathfrak{A}_A^\dag(\theta_1)_{b_1a_1}\mathfrak{A}_A^\dag(\theta_2)_{b_2a_2}|0\rangle\nonumber\\
&=&\frac{1}{N}[p_1+p_2-p_3-p_4]_\mu\,\vec{L}(\theta_1,\theta_2,\theta_3,\theta_4)\cdot \vec{D}_{a_0c_0a_1a_2a_3a_4;b_1b_2b_3b_4},\label{fourparticlefive}
\eeq
for $\theta_3>\theta_4>\theta_1>\theta_2$,
\beq
\langle 0|j_\mu^L(0)_{a_0c_0}\!\!\!\!\!&|&\!\!\!\!\! P,\theta_1,a_1,b_1;A,\theta_2,b_2,a_2;A,\theta_3,b_3,a_3;P,\theta_4,a_4,b_4\rangle\nonumber\\
&=&\langle 0|j_\mu^L(0)_{a_0c_0}\mathfrak{A}_{P}^\dag(\theta_3)_{a_3a_3}\mathfrak{A}_A^\dag(\theta_1)_{b_1a_1}\mathfrak{A}_A^\dag(\theta_2)_{b_2a_2}\mathfrak{A}_P^\dag(\theta_4)_{a_4b_4}|0\rangle\nonumber\\
&=&\frac{1}{N}[p_1+p_2-p_3-p_4]_\mu\,\vec{Q}(\theta_1,\theta_2,\theta_3,\theta_4)\cdot \vec{D}_{a_0c_0a_1a_2a_3a_4;b_1b_2b_3b_4},\label{fourparticlesix}
\eeq
for $\theta_3>\theta_1>\theta_2>\theta_4$,
\beq
\langle 0|j_\mu^L(0)_{a_0c_0}\!\!\!\!\!&|&\!\!\!\!\! A,\theta_2,b_2,a_2;A,\theta_1,b_1,a_1;P,\theta_3,a_3,b_3;P,\theta_4,a_4,b_4\rangle\nonumber\\
&=&\langle 0| j_\mu^L(0)_{a_0c_0} \mathfrak{A}_{A}^\dag(\theta_2)_{b_2a_2}\mathfrak{A}_A^\dag(\theta_1)_{b_1a_1}\mathfrak{A}_P^\dag(\theta_3)_{a_3b_3}\mathfrak{A}_P^\dag(\theta_4)_{a_4b_4}|0\rangle\nonumber\\
&=&\frac{1}{N}[p_1+p_2-p_3-p_4]_\mu\,\vec{F}(\theta_2,\theta_1,\theta_3,\theta_4)\cdot \vec{D}_{a_0c_0a_1a_2a_3a_4;b_1b_2b_3b_4},\label{fourparticleseven}
\eeq
for $\theta_2>\theta_1>\theta_3,>\theta_4$, and
\beq
\langle 0|j_\mu^L(0)_{a_0c_0}\!\!\!\!\!&|&\!\!\!\!\! A,\theta_1,b_1,a_1;A,\theta_2,b_2,a_2;P,\theta_4,a_4,b_4;P,\theta_3,a_3,b_3\rangle\nonumber\\
&=&\langle 0| j_\mu^L(0)_{a_0c_0} \mathfrak{A}_{A}^\dag(\theta_1)_{b_1a_1}\mathfrak{A}_A^\dag(\theta_2)_{b_2a_2}\mathfrak{A}_P^\dag(\theta_4)_{a_4b_4}\mathfrak{A}_P^\dag(\theta_3)_{a_3b_3}|0\rangle\nonumber\\
&=&\frac{1}{N}[p_1+p_2-p_3-p_4]_\mu\,\vec{F}(\theta_1,\theta_2,\theta_4,\theta_3)\cdot \vec{D}_{a_0c_0a_1a_2a_3a_4;b_1b_2b_3b_4},\label{fourparticleeight}
\eeq
for $\theta_1>\theta_2>\theta_4>\theta_3$,
where we define the eight-component vectors
\beq
[\vec{D}_{a_0c_0a_1a_2a_3a_4;b_1b_2b_3b_4}]=\left(\begin{array}{c}\delta_{a_0a_3}\delta_{a_1c_0}\delta_{a_2a_4}\delta_{b_1b_3}\delta_{b_2b_4}-\frac{1}{N}\delta_{a_0c_0}\delta_{a_1a_3}\delta_{a_2a_4}\delta_{b_1b_3}\delta_{b_2b_4}\\\,\\\delta_{a_0a_3}\delta_{a_1c_0}\delta_{a_2a_4}\delta_{b_1b_4}\delta_{b_2b_3}-\frac{1}{N}\delta_{a_0c_0}\delta_{a_1a_3}\delta_{a_2a_4}\delta_{b_1b_4}\delta_{b_2b_3}\\\,\\\delta_{a_0a_4}\delta_{a_1c_0}\delta_{a_2a_3}\delta_{b_1b_3}\delta_{b_2b_4}-\frac{1}{N}\delta_{a_0c_0}\delta_{a_1a_4}\delta_{a_2a_3}\delta_{b_1b_3}\delta_{b_2b_4}\\\,\\\delta_{a_0a_4}\delta_{a_1c_0}\delta_{a_2a_3}\delta_{b_1b_4}\delta_{b_2b_3}-\frac{1}{N}\delta_{a_0c_0}\delta_{a_1a_4}\delta_{a_2a_3}\delta_{b_1b_4}\delta_{b_2b_3}\\\,\\\delta_{a_0a_3}\delta_{a_1a_4}\delta_{a_2c_0}\delta_{b_1b_4}\delta_{b_2b_3}-\frac{1}{N}\delta_{a_0c_0}\delta_{a_2a_3}\delta_{a_1a_4}\delta_{b_1b_4}\delta_{b_2b_3}\\\,\\\delta_{a_0a_3}\delta_{a_1a_4}\delta_{a_2c_0}\delta_{b_1b_3}\delta_{b_2b_4}-\frac{1}{N}\delta_{a_0c_0}\delta_{a_2a_3}\delta_{a_1a_4}\delta_{b_1b_3}\delta_{b_2b_4}\\\,\\\delta_{a_0a_4}\delta_{a_1a_3}\delta_{a_2c_0}\delta_{b_1b_3}\delta_{b_2b_4}-\frac{1}{N}\delta_{a_0c_0}\delta_{a_2a_4}\delta_{a_1a_3}\delta_{b_1b_3}\delta_{b_2b_4}\\\,\\\delta_{a_0a_4}\delta_{a_1a_3}\delta_{a_2a_0}\delta_{b_1b_4}\delta_{b_2b_3}-\frac{1}{N}\delta_{a_0c_0}\delta_{a_2a_4}\delta_{a_1a_3}\delta_{b_1b_4}\delta_{b_2b_3}\end{array}\right),
\eeq
\beq
[\vec{F}(\theta_1,\theta_2,\theta_3,\theta_4)]=\left(\begin{array}{c}F_1(\theta_1,\theta_2,\theta_3,\theta_4)\\\,\\F_2(\theta_1,\theta_2,\theta_3,\theta_4)\\\,\\F_3(\theta_1,\theta_2,\theta_3,\theta_4)\\\,\\F_4(\theta_1,\theta_2,\theta_3,\theta_4)\\\,\\F_5(\theta_1,\theta_2,\theta_3,\theta_4)\\\,\\F_6(\theta_1,\theta_2,\theta_3,\theta_4)\\\,\\F_7(\theta_1,\theta_2,\theta_3,\theta_4)\\\,\\F_8(\theta_1,\theta_2,\theta_3,\theta_4)\end{array}\right),\nonumber
\eeq
and similarly for $\vec{G},\,\vec{H},\,\vec{K},\,\vec{L}$ and $\vec{Q}$.

Watson's theorem relates the form factors with different ordering of rapidities, 
yielding
\beq
\langle 0|j_\mu^L(0\!\!\!\!\!&)&\!\!\!\!\! _{a_0c_0}\mathfrak{A}_A^\dag(\theta_1)_{b_1a_1}\mathfrak{A}_P^\dag(\theta_3)_{a_3b_3}\mathfrak{A}_A^\dag(\theta_2)_{b_2a_2}\mathfrak{A}_P^\dag(\theta_4)_{a_4b_4}|0\rangle\nonumber\\
&=&S_{AP}(\theta_{23})_{a_3b_3;b_2a_2}^{d_2c_2;c_3d_3}\langle 0|j_\mu^L(0)_{a_0c_0}\mathfrak{A}_A^\dag(\theta_1)_{b_1a_1}\mathfrak{A}_A^\dag(\theta_2)_{d_2c_2}\mathfrak{A}_P^\dag(\theta_3)_{c_3d_3}\mathfrak{A}_P^\dag(\theta_4)_{a_4b_4}|0\rangle,\nonumber
\eeq
\beq
\langle 0|j_\mu^L(0\!\!\!\!\!&)&\!\!\!\!\!_{a_0c_0}\mathfrak{A}_A^\dag(\theta_1)_{b_1a_1}\mathfrak{A}_P^\dag(\theta_3)_{a_3b_3}\mathfrak{A}_P^\dag(\theta_4)_{a_4b_4}\mathfrak{A}_A^\dag(\theta_2)_{b_2a_2}|0\rangle\nonumber\\
&=&S_{AP}(\theta_{24})_{a_4b_4;b_2a_2}^{d_2c_2;c_4d_4}\langle 0|j_\mu^L(0)_{a_0c_0} \mathfrak{A}_A^\dag(\theta_1)_{b_1a_1}\mathfrak{A}_P^\dag(\theta_3)_{a_3b_3}\mathfrak{A}_A^\dag(\theta_2)_{d_2c_2}\mathfrak{A}_P^\dag(\theta_4)_{c_4d_4}|0\rangle\nonumber,
\eeq
\beq
\langle 0| j_\mu^L(0\!\!\!\!\!&)&\!\!\!\!\!_{a_0c_0}\mathfrak{A}_P^\dag(\theta_3)_{a_3b_3}\mathfrak{A}_A^\dag(\theta_1)_{b_1a_1}\mathfrak{A}_P^\dag(\theta_4)_{a_4b_4}\mathfrak{A}_A^\dag(\theta_2)_{b_2a_2}|0\rangle\nonumber\\
&=&S_{AP}(\theta_{13})^{d_1c_1;c_3d_3}_{a_3b_3;b_1a_1}\langle 0|j_\mu^L(0)_{a_0c_0}\mathfrak{A}_A^\dag(\theta_1)_{d_1c_1}\mathfrak{A}_P^\dag(\theta_3)_{c_3d_3}\mathfrak{A}_P^\dag(\theta_4)_{a_4b_4}\mathfrak{A}_A^\dag(\theta_2)_{b_2a_2}|0\rangle,\nonumber
\eeq
\beq
\langle0|j_\mu^L(0\!\!\!\!\!&)&\!\!\!\!\!_{a_0c_0}\mathfrak{A}_P^\dag(\theta_3)_{a_3b_3}\mathfrak{A}_P^\dag(\theta_4)_{a_4b_4}\mathfrak{A}_A^\dag(\theta_1)_{b_1a_1}\mathfrak{A}_A^\dag(\theta_2)_{b_2a_1}|0\rangle\nonumber\\
&=&S_{AP}(\theta_{14})^{d_1c_1;c_4d_4}_{a_4b_4;b_1a_1}\langle0|j_\mu^L(0)_{a_0c_0}\mathfrak{A}_P^\dag(\theta_3)_{a_3b_3}\mathfrak{A}_A^\dag(\theta_1)_{d_1c_1}\mathfrak{A}_P^\dag(\theta_4)_{c_4d_4}\mathfrak{A}_A^\dag(\theta_2)_{b_2a_2}|0\rangle,\nonumber
\eeq
\beq
\langle 0|j_\mu^L(0\!\!\!\!\!&)&\!\!\!\!\!_{a_0c_0}\mathfrak{A}_P^\dag(\theta_3)_{a_3b_3}\mathfrak{A}_A^\dag(\theta_1)_{b_1a_1}\mathfrak{A}_A^\dag(\theta_2)_{b_2a_2}\mathfrak{A}_P^\dag(\theta_4)_{a_4b_4}|0\rangle\nonumber\\
&=&S_{AP}(\theta_{13})^{d_1c_1;c_3d_3}_{a_3b_3;b_1a_1}\langle 0|j_\mu^L(0)_{a_0c_0}\mathfrak{A}_A^\dag(\theta_1)_{d_1c_1}\mathfrak{A}_P^\dag(\theta_3)_{c_3d_3}\mathfrak{A}_A^\dag(\theta_2)_{b_2a_2}\mathfrak{A}_P^\dag(\theta_4)_{a_4b_4}|0\rangle,\nonumber
\eeq
\beq
\langle 0|j_\mu^L(0\!\!\!\!\!&)&\!\!\!\!\!_{a_0c_0}\mathfrak{A}_A^\dag(\theta_1)_{b_1a_1}\mathfrak{A}_A^\dag(\theta_2)_{b_2a_2}\mathfrak{A}_P^\dag(\theta_3)_{a_3b_3}\mathfrak{A}_P(\theta_4)_{a_4b_4}|0\rangle\nonumber\\
&=&S_{AA}(\theta_{12})^{d_2c_2;d_1c_1}_{b_1a_1;b_2a_2}\langle0|j_\mu^L(0)_{a_0c_0}\mathfrak{A}_A^\dag(\theta_2)_{d_2c_2}\mathfrak{A}_A^\dag(\theta_1)_{d_1c_1}\mathfrak{A}_P^\dag(\theta_3)_{a_3b_3}\mathfrak{A}_P^\dag(\theta_4)_{a_4b_4}|0\rangle,\nonumber
\eeq
\beq
\langle 0|j_\mu^L(0\!\!\!\!\!&)&\!\!\!\!\!_{a_0c_0}\mathfrak{A}_A^\dag(\theta_1)_{b_1a_1}\mathfrak{A}_A^\dag(\theta_2)_{b_2a_2}\mathfrak{A}_P^\dag(\theta_3)_{a_3b_3}\mathfrak{A}_P(\theta_4)_{a_4b_4}|0\rangle\nonumber\\
&=&S_{PP}(\theta_{34})^{c_4d_4;c_3d_3}_{a_3b_3;a_4b_4}\langle0|j_\mu^L(0)_{a_0c_0}\mathfrak{A}_A^\dag(\theta_1)_{b_1a_1}\mathfrak{A}_A^\dag(\theta_2)_{b_2a_2}\mathfrak{A}_P^\dag(\theta_4)_{c_4d_4}\mathfrak{A}_P^\dag(\theta_3)_{c_3d_3}|0\rangle.\nonumber
\eeq
These imply, respectively,
\beq
\vec{G}(\theta_1,\theta_2,\!\!\!&\theta_3&\!\!\!,\theta_4)\nonumber\\
&=&\left(\begin{array}{cccccccc}
1&0&0&0&0&0&0&0\\
\frac{-2\pi i}{N\hat{\theta}_{23}}&\left(1-\frac{2\pi i}{\hat{\theta}_{23}}\right)&0&0&0&0&0&0\\
\frac{-2\pi i}{N\hat{\theta}_{23}}&0&\left(1-\frac{2\pi i}{\hat{\theta}_{23}}\right)&0&0&0&\frac{-2\pi i}{N\hat{\theta}_{23}}&0\\
0&\frac{-1}{N}\left(\frac{2\pi i}{\hat{\theta}_{23}}+\frac{4\pi^2}{\hat{\theta}_{23}^2}\right)&\frac{-1}{N}\left(\frac{2\pi i}{\hat{\theta}_{23}}+\frac{4\pi^2}{\hat{\theta}_{23}^2}\right)&\left(1-\frac{4\pi i}{\hat{\theta}_{23}}-\frac{4\pi^2}{\hat{\theta}_{23}^2}\right)&0&0&0&\frac{-1}{N}\left(\frac{2\pi i}{\hat{\theta}_{23}}+\frac{4\pi^2}{\hat{\theta}_{23}^2}\right)\\
0&0&0&0&\left(1-\frac{2\pi i}{\hat{\theta}_{23}}\right)&\frac{-2\pi i}{N\hat{\theta}_{23}}&0&0\\
0&0&0&0&0&1&0&0\\
0&0&0&0&0&0&1&0\\
0&0&0&0&0&0&\frac{-2\pi i}{N\hat{\theta}_{23}}&\left(1-\frac{2\pi i}{\hat{\theta}_{23}}\right)
\end{array}\right)\nonumber\\
&&\times\vec{F}(\theta_1,\theta_2,\theta_3,\theta_4) +\mathcal{O}\left(\frac{1}{N^2}\right)\nonumber\\
&\equiv& \overleftrightarrow{M}_1(\theta_2,\theta_3)\vec{F}(\theta_1\theta_2,\theta_3,\theta_4)+\mathcal{O}\left(\frac{1}{N^2}\right),\label{watsonone}
\eeq
\beq
\vec{H}(\theta_1,\theta_2,\!\!\!&\theta_3&\!\!\!,\theta_4)\nonumber\\
&=&\left(\begin{array}{cccccccc}
\left(1-\frac{4\pi i}{\hat{\theta}_{24}}-\frac{4\pi^2}{\hat{\theta}^2_{24}}\right)&\frac{-1}{N}\left(\frac{2\pi i}{\hat{\theta}_{24}}+\frac{4\pi^2}{\hat{\theta}^2_{24}}\right)&\frac{-1}{N}\left(\frac{2\pi i}{\hat{\theta}_{24}}+\frac{4\pi^2}{\hat{\theta}^2_{24}}\right)&0&0&\frac{-1}{N}\left(\frac{2\pi i}{\hat{\theta}_{24}}-\frac{4\pi^2}{\hat{\theta}_{24}^2}\right)&0&0\\
0&\left(1-\frac{2\pi i}{\hat{\theta}_{24}}\right)&0&\frac{-2\pi i}{N\hat{\theta}_{24}}&\frac{-2\pi i}{N\hat{\theta}_{24}^2}&0&0&0\\
0&0&\left(1-\frac{2\pi i}{\hat{\theta}_{24}}\right)&\frac{-2\pi i}{N\hat{\theta}_{24}}&0&0&0&0\\
0&0&0&1&0&0&0&0\\
0&0&0&0&1&0&0&0\\
0&0&0&0&\frac{-2\pi i}{N\hat{\theta}_{24}}&\left(1-\frac{2\pi i}{\hat{\theta}_{24}}\right)&0&0\\
0&0&0&0&0&0&\left(1-\frac{2\pi i}{\hat{\theta}_{24}}\right)&\frac{-2\pi i}{N\hat{\theta}_{24}}\\
0&0&0&0&0&0&0&1
\end{array}\right)\nonumber\\
&&\times \vec{G}(\theta_1,\theta_2,\theta_3,\theta_4)+\mathcal{O}\left(\frac{1}{N^2}\right)\nonumber\\
&\equiv&\overleftrightarrow{M}_2(\theta_2,\theta_4)\vec{G}(\theta_1,\theta_2,\theta_3,\theta_4)+\mathcal{O}\left(\frac{1}{N^2}\right),\label{watsontwo}
\eeq
\beq
\vec{K}(\theta_1,\theta_2,\!\!\!&\theta_3&\!\!\!,\theta_4)\nonumber\\
&=&\left(\begin{array}{cccccccc}
\left(1-\frac{2\pi i}{\hat{\theta}_{13}}\right)&\frac{-2\pi i}{N\hat{\theta}_{13}}&0&0&0&0&0&0\\
0&1&0&0&0&0&0&0\\
0&0&\left(1-\frac{2\pi i}{\hat{\theta}_{13}}\right)&\frac{-2\pi i}{N\hat{\theta}_{13}}&0&0&0&0\\
0&0&0&1&0&0&0&0\\
0&0&0&0&1&0&0&0\\
0&0&0&0&\frac{-2\pi i}{N\hat{\theta}_{13}}&\left(1-\frac{2\pi i}{\hat{\theta}_{13}}\right)&0&0\\
0&0&0&0&0&\frac{-1}{N}\left(\frac{2\pi i}{\hat{\theta}_{13}}+\frac{4\pi^2}{\hat{\theta}_{13}^2}\right)&\left(1-\frac{4\pi i}{\hat{\theta}_{13}}-\frac{4\pi^2}{\hat{\theta}_{13}^2}\right)&\frac{-1}{N}\left(\frac{2\pi i}{\hat{\theta}_{13}}+\frac{4\pi^2}{\hat{\theta}_{13}^2}\right)\\
0&0&0&\frac{-2\pi i}{N\hat{\theta}_{13}}&\frac{-2\pi i}{N\hat{\theta}_{13}}&0&0&\left(1-\frac{2\pi i}{\hat{\theta}_{13}}\right)
\end{array}\right)\nonumber\\
&&\times\vec{H}(\theta_1,\theta_2,\theta_3,\theta_4)+\mathcal{O}\left(\frac{1}{N^2}\right)\nonumber\\
&\equiv& \overleftrightarrow{M}_3(\theta_1,\theta_3)\vec{H}(\theta_1,\theta_2,\theta_3,\theta_4)+\mathcal{O}\left(\frac{1}{N^2}\right),\label{watsonthree}
\eeq
\beq
\vec{L}(\theta_1,\theta_2,\!\!\!&\theta_3&\!\!\!,\theta_4)\nonumber\\
&=&\left(\begin{array}{cccccccc}
1&0&0&0&0&0&0&0\\
\frac{-2\pi i}{N\hat{\theta}_{14}}&\left(1-\frac{2\pi i}{\hat{\theta}_{14}}\right)&0&0&0&0&0&0\\
0&0&1&0&0&0&0&0\\
0&0&\frac{-2\pi i}{N\hat{\theta}_{14}}&\left(1-\frac{2\pi i}{\hat{\theta}_{14}}\right)&0&0&0&0\\
0&0&0&0&\left(1-\frac{4\pi i}{\hat{\theta}_{14}}-\frac{4\pi^2}{\hat{\theta}_{14}^2}\right)&\frac{-1}{N}\left(\frac{2\pi i}{\hat{\theta}_{14}}+\frac{4\pi^2}{\hat{\theta}_{14}^2}\right)&0&\frac{-1}{N}\left(\frac{2\pi i}{\hat{\theta}_{14}}+\frac{4\pi^2}{\hat{\theta}_{14}^2}\right)\\
\frac{-2\pi i}{N\hat{\theta}_{14}}&0&0&0&0&\left(1-\frac{2\pi i}{\hat{\theta}_{14}}\right)&\frac{-2\pi i}{N\hat{\theta}_{14}}&0\\
0&0&0&0&0&0&1&0\\
0&0&0&0&0&0&\frac{-2\pi i}{N\hat{\theta}_{14}}&\left(1-\frac{2\pi i}{\hat{\theta}_{14}}\right)
\end{array}\right)\nonumber\\
&&\times\vec{K}(\theta_1,\theta_2,\theta_3,\theta_4)+\mathcal{O}\left(\frac{1}{N^2}\right)\nonumber\\
&\equiv&\overleftrightarrow{M}_4(\theta_1,\theta_4)\vec{K}(\theta_1,\theta_2,\theta_3,\theta_4)+\mathcal{O}\left(\frac{1}{N^2}\right),\label{watsonfour}
\eeq
\beq
\vec{Q}(\theta_1,\theta_2,\theta_3,\theta_4)=\overleftrightarrow{M}_3(\theta_1,\theta_3)\vec{G}(\theta_1,\theta_2,\theta_3,\theta_4)+\mathcal{O}\left(\frac{1}{N^2}\right),\label{watsonfive}
\eeq
\beq
\vec{F}(\theta_1,\theta_2,\theta_3,\theta_4)&=&\left(\begin{array}{cccccccc}
0&\frac{-2\pi i}{N{\theta}_{12}}&0&0&1&\frac{-2\pi i}{N\theta_{12}}&0&0\\
\frac{-2\pi i}{N\theta_{12}}&0&0&0&\frac{-2\pi i}{N\theta_{12}}&1&0&0\\
0&0&0&\frac{-2\pi i}{N\theta_{12}}&0&0&\frac{-2\pi i}{N\theta_{12}}&1\\
0&0&\frac{-2\pi i}{N\theta_{12}}&0&0&0&1&\frac{-2\pi i}{N\theta_{12}}\\
1&\frac{-2\pi i}{N\theta_{12}}&0&0&0&\frac{-2\pi i}{N\theta_{12}}&0&0\\
\frac{-2\pi i}{N\theta_{12}}&1&0&0&\frac{-2\pi i}{N\theta_{12}}&0&0&0\\
0&0&\frac{-2\pi i}{N\theta_{12}}&1&0&0&0&\frac{-2\pi i}{N\theta_{12}}\\
0&0&1&\frac{-2\pi i}{N\theta_{12}}&0&0&\frac{-2\pi i}{N\theta_{12}}&0
\end{array}\right)\vec{F}(\theta_2,\theta_1,\theta_3,\theta_4)+\mathcal{O}\left(\frac{1}{N^2}\right)\nonumber\\
&\equiv&\overleftrightarrow{I}_1(\theta_1,\theta_2)\vec{F}(\theta_2,\theta_1,\theta_3,\theta_4)+\mathcal{O}\left(\frac{1}{N^2}\right),\label{watsonsix}
\eeq
\beq
\vec{F}(\theta_1,\theta_2,\theta_3,\theta_4)&=&\left(\begin{array}{cccccccc}
0&\frac{-2\pi i}{N\theta_{34}}&\frac{-2\pi i}{N\theta_{34}}&1&0&0&0&0\\
\frac{-2\pi i}{N\theta_{34}}&0&1&\frac{-2\pi i}{N\theta_{34}}&0&0&0&0\\
\frac{-2\pi i}{N\theta_{34}}&1&0&\frac{-2\pi i}{N\theta_{34}}&0&0&0&0\\
1&\frac{-2\pi i}{N\theta_{34}}&\frac{-2\pi i}{N\theta_{34}}&0&0&0&0&0\\
0&0&0&0&0&\frac{-2\pi i}{N\theta_{34}}&1&\frac{-2\pi i}{N\theta_{34}}\\
0&0&0&0&\frac{-2\pi i}{N\theta_{34}}&0&\frac{-2\pi i}{N\theta_{34}}&1\\
0&0&0&0&1&\frac{-2\pi i}{N\theta_{34}}&0&\frac{-2\pi i}{N\theta_{34}}\\
0&0&0&0&\frac{-2\pi i}{N\theta_{34}}&1&\frac{-2\pi i}{N\theta_{34}}&0
\end{array}\right)\vec{F}(\theta_1,\theta_2,\theta_4,\theta_3)+\mathcal{O}\left(\frac{1}{N^2}\right)\nonumber\\
&\equiv&\overleftrightarrow{I}_2(\theta_3,\theta_4)\vec{F}(\theta_1,\theta_2,\theta_4,\theta_3)+\mathcal{O}\left(\frac{1}{N^2}\right).\label{watsonseven}
\eeq

Next we apply the Smirnov periodicity axiom (\ref{periodicity}):
\beq
\langle 0|j_\mu^L(0\!\!\!\!\!&)&\!\!\!\!\!_{a_0c_0}\mathfrak{A}_A^\dag(\theta_1-2\pi i)_{b_1a_1}\mathfrak{A}_A^\dag(\theta_2)_{b_2a_2}\mathfrak{A}_P^\dag(\theta_3)_{a_3b_3}\mathfrak{A}_P^\dag(\theta_4)_{a_4b_4}|0\rangle\nonumber\\
&=&\langle0|j_\mu^L(0)_{a_0c_0}\mathfrak{A}_A^\dag(\theta_2)_{b_2a_2}\mathfrak{A}_P^\dag(\theta_3)_{a_3b_3}\mathfrak{A}_P^\dag(\theta_4)_{a_4b_4}\mathfrak{A}_A^\dag(\theta_1)_{b_1a_1}|0\rangle,\nonumber\\
&&\,\nonumber\\
\langle 0|j_\mu^L(0\!\!\!\!\!&)&\!\!\!\!\!_{a_0c_0}\mathfrak{A}_A^\dag(\theta_2-2\pi i)_{b_2a_2}\mathfrak{A}_P^\dag(\theta_3)_{a_3b_3}\mathfrak{A}_P^\dag(\theta_4)_{a_4b_4}\mathfrak{A}_A^\dag(\theta_1)_{b_1a_1}|0\rangle\nonumber\\
&=&\langle0|j_\mu^L(0)_{a_0c_0}\mathfrak{A}_P^\dag(\theta_3)_{a_3b_3}\mathfrak{A}_P^\dag(\theta_4)_{a_4b_4}\mathfrak{A}_A^\dag(\theta_1)_{b_1a_1}\mathfrak{A}_A^\dag(\theta_2)_{b_2a_2}|0\rangle,\nonumber\\
&&\,\nonumber\\
\langle 0|j_\mu^L(0\!\!\!\!\!&)&\!\!\!\!\!_{a_0c_0}\mathfrak{A}_P^\dag(\theta_3-2\pi i)_{a_3b_3}\mathfrak{A}_P^\dag(\theta_4)_{a_4b_4}\mathfrak{A}_A^\dag(\theta_1)_{b_1a_1}\mathfrak{A}_A^\dag(\theta_2)_{b_2a_2}|0\rangle\nonumber\\
&=&\langle0|j_\mu^L(0)_{a_0c_0}\mathfrak{A}_P^\dag(\theta_4)_{a_4b_4}\mathfrak{A}_A^\dag(\theta_1)_{b_1a_1}\mathfrak{A}_A^\dag(\theta_2)_{b_2a_2}\mathfrak{A}_P^\dag(\theta_3)_{a_3b_3}|0\rangle,\nonumber\\
&&\,\nonumber\\
\langle 0|j_\mu^L(0\!\!\!\!\!&)&\!\!\!\!\!_{a_0c_0}\mathfrak{A}_P^\dag(\theta_4-2\pi i)_{a_4b_4}\mathfrak{A}_A^\dag(\theta_1)_{b_1a_1}\mathfrak{A}_A^\dag(\theta_2)_{b_2a_2}\mathfrak{A}_P^\dag(\theta_3)_{a_3b_3}|0\rangle\nonumber\\
&=&\langle0|j_\mu^L(0)_{a_0c_0}\mathfrak{A}_A^\dag(\theta_1)_{b_1a_1}\mathfrak{A}_A^\dag(\theta_2)_{b_2a_2}\mathfrak{A}_P^\dag(\theta_3)_{a_3b_3}\mathfrak{A}_P^\dag(\theta_4)_{a_4b_4}|0\rangle,\nonumber
\eeq
which imply, respectively,
\beq
\vec{F}(\theta_1-2\pi i,\theta_2,\theta_3,\theta_4)&=&\vec{H}(\theta_2,\theta_1,\theta_3,\theta_4),\label{smirnovone}\\
\vec{H}(\theta_2-2\pi i,\theta_1,\theta_3,\theta_4)&=&\vec{L}(\theta_1,\theta_2,\theta_3,\theta_4),\label{smirnovtwo}\\
\vec{L}(\theta_1,\theta_2,\theta_3-2\pi i,\theta_4)&=&\vec{Q}(\theta_1,\theta_2,\theta_4,\theta_3),\label{smirnovthree}\\
\vec{Q}(\theta_1,\theta_2,\theta_4-2\pi i,\theta_3)&=&\vec{F}(\theta_1,\theta_2,\theta_3,\theta_4).\label{smirnovfour}
\eeq

We combine Watson's theorem with the periodicity axiom, to express Equations (\ref{smirnovone}),
 (\ref{smirnovtwo}),  (\ref{smirnovthree}) and (\ref{smirnovfour})
in terms of only 
$\vec{F}(\theta_1,\theta_2,\theta_3,\theta_4)$. We combine (\ref{smirnovone}) with (\ref{watsonfour}), (\ref{watsonthree}) and (\ref{watsonsix}), and find 
\beq
\vec{F}(\theta_1-2\pi i,\theta_2,\theta_3,\theta_4)=\overleftrightarrow{M}_4(\theta_1,\theta_4)\overleftrightarrow{M}_3(\theta_1,\theta_3)\left[\overleftrightarrow{I}_1(\theta_1,\theta_2)\right]^{-1}\vec{F}(\theta_1,\theta_2,\theta_3,\theta_4).\label{watsonandsmirnovone}
\eeq
Combining (\ref{smirnovtwo}) with (\ref{watsontwo}), (\ref{watsonone}) and (\ref{watsonsix}) gives
\beq
\left[\overleftrightarrow{I}_1(\theta_1,\theta_2-2\pi i)\right]^{-1}\vec{F}(\theta_1,\theta_2-2\pi i,\theta_3,\theta_4)=\overleftrightarrow{M}_2(\theta_2,\theta_4)\overleftrightarrow{M}_1(\theta_2,\theta_4)\vec{F}(\theta_1,\theta_2,\theta_3,\theta_4).\label{watsonandsmirnovtwo}
\eeq
Combining (\ref{smirnovthree}) with (\ref{watsonthree}), (\ref{watsonone}) and (\ref{watsonseven}) gives
\beq
\overleftrightarrow{M}_3(\theta_1,\theta_3-2\pi i)\overleftrightarrow{M}_1(\theta_2,\theta_3-2\pi i)\vec{F}(\theta_1,\theta_2,\theta_3,\theta_4)=\left[\overleftrightarrow{I}_2(\theta_3,\theta_4)\right]^{-1}\vec{F}(\theta_1,\theta_2,\theta_3,\theta_4).\label{watsonandsmirnovthree}
\eeq
Finally, we combine (\ref{smirnovfour}) with (\ref{watsonfour}), (\ref{watsontwo}) and (\ref{watsonseven}) to find
\beq
\overleftrightarrow{M}_4(\theta_1,\theta_4-2\pi i)\overleftrightarrow{M}_2(\theta_2,\theta_4-2\pi i)\left[\overleftrightarrow{I}_2(\theta_3,\theta_4-2\pi i)\right]^{-1}\vec{F}(\theta_1,\theta_2,\theta_3,\theta_4-2\pi i)=\vec{F}(\theta_1,\theta_2,\theta_3,\theta_4).\label{watsonandsmirnovfour}
\eeq

The set of equations (\ref{watsonandsmirnovone}), (\ref{watsonandsmirnovtwo}), (\ref{watsonandsmirnovthree}) and (\ref{watsonandsmirnovfour}) are difficult to solve, for finite $N$. In the large-$N$ limit, the matrices $\overleftrightarrow{M}_{1,2,3,4}$ become diagonal and mutually commute, and the matrices $\overleftrightarrow{I}_{1,2}$ become their own inverses. This greatly simplifies the problem, allowing us to find the form factors. We expand the form factors in powers of $1/N$ as $\vec{F}(\theta_1,\theta_2,\theta_3,\theta_4)=\vec{F}^0(\theta_1,\theta_2,\theta_3,\theta_4)+\frac{1}{N}\vec{F}^1(\theta_1,\theta_2,\theta_3,\theta_4)+\dots$, simplifying the periodicity conditions for $\vec{F}^0(\theta_1,\theta_2,\theta_3,\theta_4)$. We combine (\ref{watsonandsmirnovone}) and (\ref{watsonandsmirnovtwo}) to get
\beq
\vec{F}^0(\theta_1-2\pi i,\theta_2-2\pi i,\theta_3,\theta_4)=\overleftrightarrow{M}_4(\theta_1,\theta_4)\overleftrightarrow{M}_3(\theta_1,\theta_3)\overleftrightarrow{M}_2(\theta_2,\theta_4)\overleftrightarrow{M}_1(\theta_2,\theta_3)\vec{F}^0(\theta_1,\theta_2,\theta_3,\theta_4),\label{watsonandsmirnovoneandtwo}
\eeq
or explicitly, in terms of the components of $\vec{F}^0(\theta_1,\theta_2,\theta_3,\theta_4)$,
\beq
F_1^0(\theta_1-2\pi i,\theta_2-2\pi i,\theta_3,\theta_4)&=&\left(\frac{\theta_{13}+\pi i}{\theta_{13}-\pi i}\right)\left(\frac{\theta_{24}+\pi i}{\theta_{24}-\pi i}\right)^2F_1^0(\theta_1,\theta_2,\theta_3,\theta_4),\nonumber\\
F_2^0(\theta_1-2\pi i,\theta_2-2\pi i,\theta_3,\theta_4)&=&\left(\frac{\theta_{14}+\pi i}{\theta_{14}-\pi i}\right)\left(\frac{\theta_{23}+\pi i}{\theta_{23}-\pi i}\right)\left(\frac{\theta_{24}+\pi i}{\theta_{24}-\pi i}\right)F_2^0(\theta_1,\theta_2,\theta_3,\theta_4),\nonumber\\
F_3^0(\theta_1-2\pi i,\theta_2-2\pi i,\theta_3,\theta_4)&=&\left(\frac{\theta_{13}+\pi i}{\theta_{13}-\pi i}\right)\left(\frac{\theta_{23}+\pi i}{\theta_{23}-\pi i}\right)\left(\frac{\theta_{24}+\pi i}{\theta_{24}-\pi i}\right)F_3^0(\theta_1,\theta_2,\theta_3,\theta_4),\nonumber\\
F_4^0(\theta_1-2\pi i,\theta_2-2\pi i,\theta_3,\theta_4)&=&\left(\frac{\theta_{14}+\pi i}{\theta_{14}-\pi i}\right)\left(\frac{\theta_{23}+\pi i}{\theta_{23}-\pi i}\right)^2F_4^0(\theta_1,\theta_2,\theta_3,\theta_4),\nonumber\\
F_5^0(\theta_1-2\pi i,\theta_2-2\pi i,\theta_3,\theta_4)&=&\left(\frac{\theta_{14}+\pi i}{\theta_{14}-\pi i}\right)^2\left(\frac{\theta_{23}+\pi i}{\theta_{23}-\pi i}\right)F_5^0(\theta_1,\theta_2,\theta_3,\theta_4),\nonumber\\
F_6^0(\theta_1-2\pi i,\theta_2-2\pi i,\theta_3,\theta_4)&=&\left(\frac{\theta_{14}+\pi i}{\theta_{14}-\pi i}\right)\left(\frac{\theta_{13}+\pi i}{\theta_{13}-\pi i}\right)\left(\frac{\theta_{24}+\pi i}{\theta_{24}-\pi i}\right)F_6^0(\theta_1,\theta_2,\theta_3,\theta_4),\nonumber\\
F_7^0(\theta_1-2\pi i,\theta_2-2\pi i,\theta_3,\theta_4)&=&\left(\frac{\theta_{13}+\pi i}{\theta_{13}-\pi i}\right)^2\left(\frac{\theta_{24}+\pi i}{\theta_{24}-\pi i}\right)F_7^0(\theta_1,\theta_2,\theta_3,\theta_4),\nonumber\\
F_8^0(\theta_1-2\pi i,\theta_2-2\pi i,\theta_3,\theta_4)&=&\left(\frac{\theta_{14}+\pi i}{\theta_{14}-\pi i}\right)\left(\frac{\theta_{13}+\pi i}{\theta_{13}-\pi i}\right)\left(\frac{\theta_{23}+\pi i}{\theta_{23}-\pi i}\right)F_8^0(\theta_1,\theta_2,\theta_3,\theta_4).\nonumber
\eeq
The solution that satisfies (\ref{watsonandsmirnovoneandtwo}), (\ref{watsonsix}) and (\ref{watsonseven}) is
\beq
F_1^0(\theta_1,\theta_2,\theta_3,\theta_4)&=&\frac{g_1(\theta_1,\theta_2,\theta_3,\theta_4)}{(\theta_{13}+\pi i)(\theta_{24}+\pi i)^2},\nonumber\\
F_2^0(\theta_1,\theta_2,\theta_3,\theta_4)&=&\frac{g_2(\theta_1,\theta_2,\theta_3,\theta_4)}{(\theta_{14}+\pi i)(\theta_{23}+\pi i)(\theta_{24}+\pi i)},\nonumber\\
F_3^0(\theta_1,\theta_2,\theta_3,\theta_4)&=&\frac{g_2(\theta_1,\theta_2,\theta_4,\theta_3)}{(\theta_{13}+\pi i)(\theta_{23}+\pi i)(\theta_{24}+\pi i)},\nonumber\\
F_4^0(\theta_1,\theta_2,\theta_3,\theta_4)&=&\frac{g_1(\theta_1,\theta_2,\theta_4,\theta_3)}{(\theta_{14}+\pi i)(\theta_{23}+\pi i)^2},\nonumber\\
F_5^0(\theta_1,\theta_2,\theta_3,\theta_4)&=&\frac{g_1(\theta_2,\theta_1,\theta_3,\theta_4)}{(\theta_{14}+\pi i)^2(\theta_{23}+\pi i)},\nonumber\\
F_6^0(\theta_1,\theta_2,\theta_3,\theta_4)&=&\frac{g_2(\theta_2,\theta_1,\theta_3,\theta_4)}{(\theta_{14}+\pi i)(\theta_{13}+\pi i)(\theta_{24}+\pi i)},\nonumber\\
F_7^0(\theta_1,\theta_2,\theta_3,\theta_4)&=&\frac{g_1(\theta_2,\theta_1,\theta_4,\theta_3)}{(\theta_{13}+\pi i)^2(\theta_{24}+\pi i)},\nonumber\\
F_8^0(\theta_1,\theta_2,\theta_3,\theta_4)&=&\frac{g_2(\theta_2,\theta_1,\theta_4,\theta_3)}{(\theta_{14}+\pi i)(\theta_{13}+\pi i)(\theta_{23}+\pi i)},\label{solutionwatsonandsmirnovoneandtwo}
\eeq
where the functions $g_1(\theta_1,\theta_2,\theta_3,\theta_4)$ and $g_2(\theta_1,\theta_2,\theta_3,\theta_4)$ are periodic under $\theta_{1,2}\to\theta_{1,2}-2\pi i$.

Instead of the analyis of the previous paragraph, we could have combined (\ref{watsonandsmirnovthree}) and (\ref{watsonandsmirnovfour}) to obtain
\beq
\overleftrightarrow{M}_4(\theta_1,\theta_4-2\pi i\!\!\!\!&)&\!\!\!\!\overleftrightarrow{M}_3(\theta_1,\theta_3-2\pi i)\overleftrightarrow{M}_2(\theta_2,\theta_4-2\pi i)\overleftrightarrow{M}_1(\theta_2,\theta_3-2\pi i)\vec{F}^0(\theta_1,\theta_2,\theta_3-2\pi i,\theta_4-2\pi i)\nonumber\\
&=&\vec{F}^0(\theta_1,\theta_2,\theta_3,\theta_4).\label{watsonandsmirnovthreeandfour}
\eeq
The condition (\ref{watsonandsmirnovthreeandfour}) is equivalent to (\ref{watsonandsmirnovoneandtwo}). The solution of (\ref{watsonandsmirnovthreeandfour}) is 
(\ref{solutionwatsonandsmirnovoneandtwo})

The minimal choice for the functions $g_{1,2}(\theta_1,\theta_2,\theta_3,\theta_4)$ is to set them equal to constants, $g_1(\theta_1,\theta_2,\theta_3,\theta_4)=g_1,\,g_2(\theta_1,\theta_2,\theta_3,\theta_4)=g_2$. These constants are fixed using the annihilation-pole axiom.
There is an annihilation pole at $\theta_{24}=-\pi i$. The annihilation-pole axiom (Equation (\ref{annihilationpoleaxiom})) implies
\beq
{\rm Res}|_{\theta_{24}=-\pi i}\langle0\!\!\!\!\!&|&\!\!\!\!\!\mathcal{O}_{a_0c_0}\mathfrak{A}_A^\dag(\theta_1)_{b_1a_1}\mathfrak{A}_P^\dag(\theta_3)_{a_3b_3}\mathfrak{A}_A^\dag(\theta_2)_{b_2a_2}\mathfrak{A}_P^\dag(\theta_4)_{a_4b_4}|0\rangle\nonumber\\
&=&2i\left\{\langle0|\mathcal{O}_{a_0c_0}\mathfrak{A}_A^\dag(\theta_1)_{b_1a_1}\mathfrak{A}_P^\dag(\theta_3)_{a_3b_3}|0\rangle\delta_{a_2a_4}\delta_{b_2b_4}\right.\nonumber\\
&&-\langle0|\mathcal{O}_{a_0c_0}\mathfrak{A}_A^\dag(\theta_1)_{b_1'a_1'}\mathfrak{A}_P(\theta_3)_{a_3'b_3'}|0\rangle\left.\delta_{a_2'a_4}\delta_{b_2'b_4}S_{AA}(\theta_{12})^{b_2'a_2';b_1'a_1'}_{d_1c_1;b_1a_1}S_{AP}(\theta_{23})^{d_1c_1;a_3'b_3'}_{a_3b_3;b_2a_2}\right\}.\label{annihilationfourparticles}
\eeq
We substitute (\ref{solutionplanar}) into the right-hand side of 
(\ref{annihilationfourparticles}) to find
\beq
\langle0|\mathcal{O}_{a_0c_0}\mathfrak{A}_A^\dag(\theta_1\!\!\!\!&)&\!\!\!\!_{b_1a_1}\mathfrak{A}_P^\dag(\theta_3)_{a_3b_3}|0\rangle\delta_{a_2a_4}\delta_{b_2b_4}\nonumber\\
-\langle0|\mathcal{O}_{a_0c_0}\mathfrak{A}_A^\dag(\theta_1\!\!\!\!&)&\!\!\!\!_{b_1'a_1'}\mathfrak{A}_P(\theta_3)_{a_3'b_3'}|0\rangle\delta_{a_2'a_4}\delta_{b_2'b_4}S_{AA}(\theta_{12})^{b_2'a_2';b_1'a_1'}_{d_1c_1;b_1a_1}S_{AP}(\theta_{23})^{d_1c_1;a_3'b_3'}_{a_3b_3;b_2a_2}\nonumber\\
&=&\frac{2\pi}{(\theta_{13}+\pi i)}\left\{\frac{2\pi i}{N\hat{\theta}_{23}}\left(\delta_{a_0a_4}\delta_{a_2a_3}\delta_{c_0a_1}\delta_{b_1b_3}\delta_{b_2b_4}-\frac{1}{N}\delta_{a_0c_0}\delta_{a_1a_4}\delta_{a_2a_3}\delta_{b_1b_3}\delta_{b_2b_4}\right)\right.\nonumber\\
&&+\frac{1}{N}\left(\frac{-2\pi i}{\hat{\theta}_{23}}+\frac{2\pi i}{\theta_{12}}-\frac{4\pi^2}{\theta_{12}\hat{\theta}_{23}}\right)\left(\delta_{a_0a_3}\delta_{a_2a_4}\delta_{a_1c_0}\delta_{b_2b_3}\delta_{b_1b_4}-\frac{1}{N}\delta_{a_0c_0}\delta_{a_1a_3}\delta_{a_2a_4}\delta_{b_2b_3}\delta_{b_1b_4}\right)\nonumber\\
&&\left.\frac{-2\pi i}{N\theta_{12}}\left(\delta_{a_0a_3}\delta_{a_1a_4}\delta_{a_2c_0}\delta_{b_1b_3}\delta_{b_2b_4}-\frac{1}{N}\delta_{a_0c_0}\delta_{a_2a_3}\delta_{a_1a_4}\delta_{b_1b_3}\delta_{b_2b_4}\right)\right\}.\nonumber
\eeq
Equation (\ref{annihilationfourparticles}) yields for the constants $g_2=8\pi^2 i$, $g_1=0$. We notice that the double poles present in (\ref{solutionwatsonandsmirnovoneandtwo}) vanish, because $g_1=0$. The first term on the right-hand side of (\ref{annihilationfourparticles}) is of order 
$1/N$. This is the reason we introduced a factor of $1/N$ in Equations (\ref{fourparticleone}) through (\ref{fourparticleeight}).

The minimal four-particle form factor satisfying all of Smirnov's axioms for large $N$ is 
\beq
\langle 0|j_\mu^L(0\!\!\!\!&)&\!\!\!\!_{a_0c_0}|A,\theta_1,b_1,a_1;A,\theta_2,b_2,a_2;P,\theta_3,a_3,b_3;P,\theta_4,a_4,b_4\rangle\nonumber\\
&=&\left[p_1+p_2-p_3-p_4\right]_\mu\frac{8\pi^2 i}{N}\nonumber\\
&&\times\left\{\frac{1}{(\theta_{14}+\pi i)(\theta_{23}+\pi i)(\theta_{24}+\pi i)}\left(\delta_{a_0a_3}\delta_{a_1c_0}\delta_{a_2a_4}\delta_{b_1b_4}\delta_{b_2b_3}-\frac{1}{N}\delta_{a_0c_0}\delta_{a_1a_3}\delta_{a_2a_4}\delta_{b_1b_4}\delta_{b_2b_3}\right)\right.\nonumber\\
&&+\frac{1}{(\theta_{13}+\pi i)(\theta_{23}+\pi i)(\theta_{24}+\pi i)}\left(\delta_{a_0a_4}\delta_{a_1c_0}\delta_{a_2a_3}\delta_{b_1b_3}\delta_{b_2b_4}-\frac{1}{N}\delta_{a_0c_0}\delta_{a_1a_4}\delta_{a_2a_3}\delta_{b_1b_3}\delta_{b_2b_4}\right)\nonumber\\
&&+\frac{1}{(\theta_{14}+\pi i)(\theta_{13}+\pi i)(\theta_{24}+\pi i)}\left(\delta_{a_0a_3}\delta_{a_1a_4}\delta_{a_2c_0}\delta_{b_1b_3}\delta_{b_2b_4}-\frac{1}{N}\delta_{a_0c_0}\delta_{a_2a_3}\delta_{a_1a_4}\delta_{b_1b_3}\delta_{b_2b_4}\right)\nonumber\\
&&\left.+\frac{1}{(\theta_{14}+\pi i)(\theta_{13}+\pi i)(\theta_{23}+\pi i)}\left(\delta_{a_0a_4}\delta_{a_1a_3}\delta_{a_2c_0}\delta_{b_1b_4}\delta_{b_2b_3}-\frac{1}{N}\delta_{a_0c_0}\delta_{a_2a_4}\delta_{a_1a_3}\delta_{b_1b_4}\delta_{b_2b_3}\right)\right\},\label{finalanswerfourparticle}
\eeq
which is the main result of this section.

\section{Conclusions}

We found the two-particle form factor of the principal-chiral-model current operator, for general $N$. 
We were only able to find the four-particle form factor for large $N$, because 
the S matrix is much simpler in this limit. 

Form factors of more excitations can be calculated at large $N$, using this method. As we add particles, the number of functions to determine grows very fast. This will be tedious, but perhaps not impossible. We hope it is possible to calculate all the form factors in the planar limit. We could use this to find Green's functions and compare with perturbation theory.

We are interested in applying the form factors found here to (2+1)-dimensional anisotropic 
Yang-Mills theory. This is a theory were the coupling constants are weak, but different in different directions. The form factors of the $O(4)$-symmetric sigma model were used to calculate the string tension \cite{orlandsix}, and the glueball masses \cite{orlandseven} of the $SU(2)$ gauge theory. We can apply our results to extend this treatment beyond the $SU(2)$ gauge group. 

\begin{acknowledgements}
I wish to thank my advisor Peter Orland for all his suggestions and helpful discussions. This project was supported in part by the National Science Foundation, under Grant No. PHY0855387.
\end{acknowledgements}

\end{document}